\def\tr{{\text{tr}}\,}

\def\epsilonF{\epsilon_{\text{F}}}
\def\kF{k_{\text{F}}}
\def\vF{v_{\text{F}}}

\def\me{m_{\text{e}}}
\def\lambdae{\lambda_{\text{e}}}
\def\muB{\mu_{\text{B}}}

\def\gso{g_{\text{so}}}
\def\be{\begin{equation}}
\def\ee{\end{equation}}
\def\bea{\begin{eqnarray}}
\def\eea{\end{eqnarray}}
\def\bse{\begin{subequations}}
\def\ese{\end{subequations}}

\documentclass[prb,twocolumn,showpacs,amsmath,amssymb,eqsecnum,preprintnumbers]{revtex4}
\usepackage{graphicx}% Include figure files
\usepackage{dcolumn}% Align table columns on decimal point
\usepackage{bm}% bold math
\begin{document}
%\preprint{Phys. Rev. B {\bf 74}, 024409 (2006)}
\title{Quantum Electrodynamics and the Origins of the
       Exchange, Dipole-Dipole, and Dzyaloshinsky-Moriya Interactions in Itinerant
       Fermion Systems}
\author{D. Belitz}
\affiliation{Department of Physics, and Institute of Theoretical Science,
             University of Oregon, Eugene, OR 97403, USA
             }
\author{T.R. Kirkpatrick}
\affiliation{Institute for Physical Science and Technology,
             and Department of Physics, University of Maryland, College Park,
             MD 20742, USA
             }
\date{\today}

\begin{abstract}
It is shown how the exchange interaction, the dipole-dipole interaction, and
the Dzyaloshinsky-Moriya interaction between electronic spin-density
fluctuations emerge naturally from a field-theoretic framework that couples
electrons to the fluctuating electromagnetic potential. Semi-quantitative
estimates are given to determine when the dipole-dipole interaction, which is
often neglected, needs to be considered, and various applications are
discussed, with an emphasis on weak ferromagnets and on helimagnets.
\end{abstract}


\maketitle

\section{Introduction}
\label{sec:I}

Understanding the origin of ferromagnetism was one of the success stories of
applying quantum mechanics to solid-state systems. Classically, magnetic
moments interact via the dipole-dipole interaction, which is much too weak to
explain magnetic order at as high a temperature as is observed in, e.g., iron
or nickel.\cite{Ashcroft_Mermin_1976} The explanation of this conundrum was
found to be the exchange interaction mechanism, which leads to a spin-spin
interaction that is governed by the Coulomb interaction via the Pauli
principle. This was first understood in the context of atomic and molecular
physics in the 1920s, and applied to solid-state physics in the
1950s.\cite{Ashcroft_Mermin_1976} Somewhat ironically, a straightforward
application of the exchange interaction concept leads to a spin-spin
interaction that is too strong, as the relevant energy scale is the atomic
scale, or roughly $100,000\,{\text K}$. Many-body and band-structure effects
renormalize this scale and bring it down to the observed ferromagnetic scale of
rougly $1,000\,{\text{K}}$ or lower.\cite{Lonzarich_Taillefer_1985} This is
still much larger than the dipole-dipole scale, and the latter is often
neglected in the discussion of ferromagnets. When it is considered, e.g., for
its influence on the critical behavior,\cite{Aharony_Fisher_1973,
Frey_Schwabl_1994, Ma_1976} it is usually added phenomenologically to models
that describe the exchange interaction. Another spin-spin interaction that has
been of interest lately is the Dzyaloshinsky-Moriya (DM)
interaction.\cite{Dzyaloshinsky_1958, Moriya_1960} It results (in systems with
suitable lattice structures) from the spin-orbit interaction, has been derived
from microscopic models, and is believed to be responsible for the helical
magnetic order observed in MnSi and FeGe.\cite{Bak_Jensen_1980} Rough estimates
show that the DM interaction and the dipole-dipole interaction are of about the
same strength, and should thus be considered together.\cite{Maleyev_2006}
Furthermore, in weak ferromagnets, which order only at low temperatures, all
three interactions can be comparable in strength, which can make the
dipole-dipole and DM interactions crucial.

In this paper we provide a comprehensive derivation of all of these effects
within one unified framework, namely, a field-theoretic description of
electrons and photons. Starting with finite-temperature quantum electrodynamics
(QED) coupled to a field-theoretic description of finite-density
quasi-relativistic electrons we show that the exchange, dipole-dipole, and DM
interactions all appear naturally upon integrating out the photons. The
exchange and DM interactions arise from integrating out the scalar part of the
electromagnetic potential; the dipole-dipole interaction, from integrating out
the vector potential. Furthermore, the DM and dipole-dipole interactions are
indeed of the same order in the relativistic corrections to the Schr{\"o}dinger
equation (i.e., of second order in $\vF/c$, with $\vF$ the Fermi velocity and
$c$ the speed of light, or of second order in the fine structure constant
$\alpha$).

Integrating out the fermions then leads to an effective theory for quantum
magnets that generalizes and replaces the Hertz-Millis theory\cite{Hertz_1976,
Millis_1993} and its generalizations.\cite{Belitz_Kirkpatrick_Vojta_2005} More
generally, the theory provides a derivation of spin-spin interactions in
itinerant Fermi systems in general, whether or not they are in a parameter
regime where they develop long-range magnetic order. Our results are therefore
relevant, for instance, for fermionic atoms in optical traps or on optical
lattices.\cite{Fregoso_Fradkin_2010}

This paper is organized as follows. In Sec.\ \ref{sec:II} we consider, as a
warm-up and to introduce various concepts, classical magnets, and show how the
vector potential coupling to the magnetization gives rise to the dipole-dipole
interaction. In Sec.\ \ref{sec:III} we develop the technical machinery for
dealing with quantum magnets and provide the derivations mentioned above. In
Sec.\ \ref{sec:IV} we discuss our results and provide a summary and conclusion.
Some technical details are relegated to various appendices.

\section{Effective Action for Classical Ferromagnets and Helimagnets}
\label{sec:II}

We now proceed to derive an effective action for magnets that includes the
effects of the fluctuating electromagnetic potential. We first consider the
classical case as a warm-up; we will generalize to the quantum case in Sec.\
\ref{sec:III}.

\subsection{Dipole-dipole interaction}
\label{subsubsec:II.A}

Consider a classical model for a ferromagnet with a three-component order
parameter ${\bm M}$. In addition to the field ${\bm M}({\bm x})$ we need to
consider the electromagnetic vector potential ${\bm A}({\bm x})$, and the
partition function $Z$ is given by
\bse
\label{eqs:2.1}
\bea
Z &=& \int D[{\bm M},{\bm A}]\ e^{S[{\bm M},{\bm A}]}
\label{eq:2.1a}\\
  &\equiv& \int D[{\bm M}]\ e^{-{\cal F}[{\bm M}]/T}.
\label{eq:2.1b}
\eea
\ese
The model is defined by specifying the action $S$, and in Eq.\ (\ref{eq:2.1b})
we have anticipated integrating out the vector potential to obtain an effective
action ${\cal F}$ in terms of the order parameter only. $T$ denotes the
temperature, so ${\cal F}$ is the free energy in mean-field approximation.
Throughout this paper we will use units such that Boltzmann's constant and
Planck's constant are equal to unity, $k_{\text{B}} = \hbar = 1$.

For the order-parameter part of $S$, we consider an $O(3)$-symmetric
$\phi^4$-theory,
\bse
\label{eqs:2.2}
\be
S_M = \frac{-1}{T}\int_V d{\bm x}\ \left[\frac{t}{2}\,{\bm M}^2({\bm x}) +
\frac{a}{2}\,(\nabla{\bm M}({\bm x}))^2 + \frac{u}{4}\,{\bm M}^4({\bm x})
\right].
\label{eq:2.2a}
\ee
$S_M$ represents a Landau-Ginzburg-Wilson (LGW) theory of an isotropic
ferromagnet with volume $V \to \infty$. The parameter $t$ contains the exchange
interaction that leads to a magnetic ordering transition. In mean-field
approximation this transition occurs at $t=0$, with $t>0$ describing the
paramagnetic phase, and $t<0$ the ferromagnetically ordered one. $a>0$ and
$u>0$ are two additional model parameters, and $(\nabla{\bm M})^2 =
\partial_i\,M_j\,\partial^i\,M^j$. Here, and throughout the paper,
summation over repeated vector, tensor, and spinor indices is implied unless
otherwise noted. Note that $S_M$ is separately invariant under rotations in
${\bm M}$ (spin) space and real space, respectively.

The magnetization\cite{magnetization_footnote} ${\bm M}$ couples linearly to
the curl of the magnetic vector potential ${\bm A}$:
\be
S_{\text{c}} = \frac{\muB}{T} \int_V d{\bm x}\ {\bm M}({\bm
x})\cdot({\bm\nabla}\times{\bm A}({\bm x})),
\label{eq:2.2b}
\ee
with $\muB = e/2\me c$ the Bohr magneton in terms of the electron charge $e$,
the electron mass $\me$, and the speed of light $c$. ${\bm A}$ and
${\bm\nabla}\times{\bm A}$ transform as vectors in real space, and therefore
$S_{\text{c}}$ is invariant only under co-rotations of spin space and real
space. It is this coupling of the magnetization to the fluctuating vector
potential that allows one to consider the magnetization as having a particular
direction in real space. The vector potential is governed by
\be
S_A = \frac{-1}{8\pi T}\int_V d{\bm x}\ \left[({\bm\nabla}\times{\bm A}({\bm
x}))^2 + \frac{1}{\rho}\,({\bm\nabla}\cdot{\bm A}({\bm x}))^2\right],
\label{eq:2.2c}
\ee
\ese
with $\rho$ any real number. The first term in Eq.\ (\ref{eq:2.2c}) is the
magnetic energy, and the second term with coupling constant $1/\rho$ is a gauge
fixing term. One popular choice is $\rho = 0$, which enforces a Coulomb gauge,
${\bm\nabla}\cdot{\bm A} = 0$; another one is the Feynman gauge,
$\rho=1$.\cite{Ryder_1985, gauge_fixing_footnote} Either choice ensures a
finite ${\bm A}$-propagator. In Coulomb gauge, it is
\be
\langle A_i({\bm k}) A_j(-{\bm k})\rangle = 4\pi T\,\frac{\delta_{ij} - {\hat
k}_i {\hat k}_j}{k^2}\ .
\label{eq:2.3}
\ee
The vector potential can now be integrated out exactly, which leads to an
effective action in terms of ${\bm M}$ only. Alternatively, we can consider the
magnetic induction ${\bm B} = {\bm\nabla}\times{\bm A}$ the fundamental field
to be integrated out. In that case, the gauge fixing condition needs to be
replaced by a constraint that enforces the Maxwell equation
${\bm\nabla}\cdot{\bm B} = 0$. That is, the Eqs.\ (\ref{eq:2.2b}),
(\ref{eq:2.2c}), and (\ref{eq:2.3}) are replaced by\cite{monopoles_footnote}
\be
S_{\text{c}} = \frac{\muB}{T} \int_V d{\bm x}\ {\bm M}({\bm x})\cdot{\bm
B}({\bm x}), \tag{2.2b$'$}
\label{eq:2.2b'}
\ee
\be
S_A = \frac{-1}{8\pi T}\int_V d{\bm x}\ \left[{\bm B}^2({\bm x}) +
\frac{1}{\rho}\,({\bm\nabla}\cdot{\bm B}({\bm x}))^2\right]_{\rho\to 0},
\tag{2.2c$'$}
\label{eq:2.2c'}
\ee
\be
\langle B_i({\bm k}) B_j(-{\bm k})\rangle = 4\pi T\,\left(\delta_{ij} - {\hat
k}_i {\hat k}_j\right). \tag{2.3$'$}
\label{eq:2.3'}
\ee
Either way we find
\bse
\label{eqs:2.4}
\bea
{\cal F} &=& \int_V d{\bm x}\ \left[\frac{1}{2}\left(t -
4\pi\,\muB^2\right)\,{\bm M}^2({\bm x}) + \frac{a}{2}\,\left(\nabla{\bm M}({\bm
x})\right)^2\right.
\nonumber\\
&&\hskip -20pt + \frac{u}{4}\,{\bm M}^4({\bm x})\biggr] + 2\pi\muB^2 \sum_{\bm
k} d_{ij}({\bm k})\,M_i({\bm k})\,M_j(-{\bm k}),
\label{eq:2.4a}
\eea
with
\be
d_{ij}({\bm k}) = {\hat k}_i {\hat k}_j\ .
\label{eq:2.4b}
\ee
\ese
The terms generated by integrating out the vector potential we recognize as the
leading contribution to the dipole-dipole interaction\cite{Aharony_Fisher_1973,
Ma_1976} plus a shift of the Landau parameter $t$ by $4\pi\muB^2$. The scalar
potential $\varphi({\bm x})$, whose gradient is the electric field, does not
lead to any magnetic interactions in a classical theory. This changes once the
system is treated quantum mechanically, see Sec.\ \ref{sec:III} below.

\subsection{Renormalization, and higher order terms}
\label{subsec:II.B}

The dipole-dipole operator $d_{ij}$, Eq.\ (\ref{eq:2.4b}), transforms as a
rank-two tensor in momentum (or real) space, and ${\bm M}$ transforms as a
vector in spin space. Consequently, the dipole-dipole interaction is invariant
under co-rotations in real space and spin space. This raises the question of
other terms in the action that have the same symmetry properties. For instance,
$({\bm\nabla}\cdot{\bm M})^2$ is allowed by symmetry. This term, and terms of
higher order in the gradient, are generated by a renormalization of the action
${\cal F}$, as we now proceed to show.

\begin{figure}[t]
\vskip -0mm
\includegraphics[width=3.0cm]{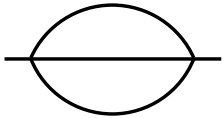}
\caption{A diagram that generates a $({\bm\nabla}\cdot{\bm M})^2$ term.}
\label{fig:2.1}
\end{figure}

A renormalization of the action ${\cal F}$ generates additional terms by virtue
of the anisotropic $M$-propagator, which now reads
\be
\langle M_i({\bm k})\,M_j(-{\bm k})\rangle = \frac{\delta_{ij} - {\hat k}_i
{\hat k}_j}{t - 4\pi\muB^2 + ak^2} + \frac{{\hat k}_i {\hat k}_j}{t + ak^2}\ .
\label{eq:2.5}
\ee
For instance, two-loop diagrams of the structure shown in Fig.\ \ref{fig:2.1}
both renormalize $d_{ij}({\bm k})$ and lead to a new vertex
\be
\sum_{\bm k} k_i\,k_j\,M_i({\bm k})\,M_j(-{\bm k}),
\label{eq:2.6}
\ee
as well as to higher order anisotropic gradient terms. Equation (\ref{eq:2.6})
represents the $({\bm\nabla}\cdot{\bm M})^2$ term that was mentioned above.
Conversely, if one starts with a theory that contains a $({\bm\nabla}\cdot{\bm
M})^2$ term, which is allowed by symmetry and hence should be included in any
Landau theory, then the leading dipole-dipole term will be generated in
perturbation theory even if it was not included in the bare action. The
complete LGW action for a classical, isotropic Heisenberg ferromagnet, up to
terms quadratic in gradients and quartic in the order parameter, thus reads
\bse
\label{eqs:2.7}
\bea
{\cal F} &=& \int_V d{\bm x}\ \left[\frac{r}{2}\,{\bm M}^2({\bm x}) +
\frac{a}{2}\,\left(\nabla{\bm M}({\bm x})\right)^2 + \frac{u}{4}\,{\bm
M}^4({\bm x})\right]
\nonumber\\
&&\hskip 0pt - \frac{d_0}{2} \int_{V} d{\bm x}\,d{\bm y}\ M_i({\bm x})\,
d_{ij}({\bm x}-{\bm y})\, M_j({\bm y})
\nonumber\\
&&\hskip 0pt + \frac{d_2}{2} \int_V d{\bm x}\ \left({\bm\nabla}\cdot{\bm
M}({\bm x})\right)^2.
\label{eq:2.7a}
\eea
Here $d_{ij}({\bm x}-{\bm y})$ is the Fourier transform of $d_{ij}({\bm k})$ in
Eq.\ (\ref{eq:2.4b}), namely,
\be
d_{ij}({\bm x}-{\bm y}) = \frac{\partial^2}{\partial x_i\,\partial x_j}\
\frac{1}{\vert {\bm x} - {\bm y} \vert}\ .
\label{eq:2.7b}
\ee
\ese
$r$ is the bare distance from the critical point in the effective LGW theory
that takes into account the effect of the vector potential, and $a$, $u$,
$d_0$, and $d_2$ are the remaining Landau parameters. As is clear from the
above discussion, one expects the bare values of $d_0$ and $d_2$ to be small of
order $1/c^2$ compared to the other parameters in natural units. These two
parameters are usually set equal to zero in elementary treatments of classical
Heisenberg ferromagnets.

\subsection{Dzyaloshinsky-Moriya interaction, and helimagnets}
\label{subsubsec:II.C}

The terms in the action so far are all even in the gradient operator, and hence
invariant under spatial inversion. The spin-orbit interaction can eliminate
this requirement by coupling the electron spins to the underlying lattice,
provided the crystal structure is not inversion invariant.
Dzyaloshinsky\cite{Dzyaloshinsky_1958} and Moriya\cite{Moriya_1960} (DM) showed
that to linear order in the spin-orbit interaction $\gso$ the relevant term is
\be
{\cal F}_{\text{DM}} = \frac{-c_{\,\text{DM}}}{2T} \int d{\bm x}\ {\bm M}({\bm
x})\cdot \left( {\bm\nabla}\times{\bm M}({\bm x})\right),
\label{eq:2.8}
\ee
with $c_{\,\text{DM}} \propto \gso$. At a classical level, this term is purely
phenomenological. DM showed how to derive it in the context of quantum
mechanics, and in Sec.\ \ref{sec:III} we will see how it arises in the context
of field theory.

\section{Quantum Systems}
\label{sec:III}

We now turn to a quantum mechanical description of itinerant fermion systems in
general, and certain types of magnetism in particular. We will show how the
exchange interaction, the dipole-dipole interaction, and the DM interaction
naturally arise in the context of a field-theoretic description of itinerant
electrons. The former two lead to ferromagnetism, and the latter, if it is
present, to helimagnetism.

\subsection{Action}
\label{subsec:III.A}

In Appendix \ref{app:A} we list the complete action for free quasi-relativistic
electrons, to order $1/c^2$, coupled to the electromagnetic field. Several
terms in the complete action are not relevant for our present purposes. The
only effect of the Darwin term, Eq.\ (\ref{eq:A.4g}), is to modify the Coulomb
interaction on length scales given by the electronic Compton wave length
$\lambdae = 1/\me c$. The relativistic mass enhancement, the last term in Eq.\
(\ref{eq:A.4b}), is a higher-order gradient term that is small compared to
terms of the same form that are generated by renormalizing the final effective
action. Finally, the Landau diamagnetic terms, Eq.\ (\ref{eq:A.4f}), give rise
to diamagnetism and, in the presence of an external magnetic field, Landau
levels. These effects are physically very different from ferromagnetism or
helimagnetism, and we do not consider them here. Finally, in quantum
electrodynamics the Fadeev-Popov ghost field does not couple to any other
fields. Its only effect is to subtract the contribution of the unphysical
longitudinal photon polarization to the free energy, and one has to keep it
only if one is interested in the absolute value of the latter. Neglecting all
of these terms, we thus consider the following action:
\bea
S[{\bar\psi},\psi;A_{\mu}] &=& \int dx\ {\bar\psi}_{\sigma}(x)
\left[-\partial_{\tau} + \frac{1}{2\me}\,{\bm\nabla}^2 + \mu
\right]\,\psi_{\sigma}(x)
\nonumber\\
&&\hskip -60pt + \frac{1}{8\pi} \int dx\
A_{\mu}(x)\,\left[\frac{1}{c^2}\,\partial_{\tau}^2 +
{\bm\nabla}^2\right]\,A_{\mu}(x) \nonumber\\
&&\hskip -60pt - ie \int dx\ \varphi(x)\,n(x) + \muB \int dx\ {\bm
B}(x)\cdot{\bm n}_{\text{s}}(x)
\nonumber\\
&&\hskip -60pt + \frac{-e}{4\me^2 c^2} \int dx\
{\bar\psi}_{\sigma_1}(x)\,{\bm\sigma}_{\sigma_1
\sigma_2}\cdot({\bm\nabla}\varphi(x)\times{\bm\nabla})\, {\psi}_{\sigma_2}(x).
\nonumber\\
\label{eq:3.1}
\eea
Here $x \equiv ({\bm x},\tau)$ comprises real space position ${\bm x}$ and
imaginary time $\tau$, and $\int dx \equiv \int_V d{\bm x} \int_0^{1/T} d\tau$.
$\bar\psi_{\sigma}$ and $\psi_{\sigma}$ are Grassmann-valued fields for
electrons with spin projection $\sigma$, and the first term in Eq.\
(\ref{eq:3.1}) describes free electrons with chemical potential
$\mu$.\cite{Negele_Orland_1988} $A_{\mu} \equiv (\phi,-A_i)$ ($\mu = 0,1,2,3$;
$i=1,2,3$) denotes the electromagnetic potential, with $\phi$ the scalar
potential and ${\bm A} = (A_1,A_2,A_3)$ the vector potential, and the second
term in in Eq.\ (\ref{eq:3.1}) describes the free electromagnetic field in
Feynman gauge (i.e., $\rho=1$ in Eq.\ (\ref{eq:A.4c})).\cite{Kaputsa_1989} Note
that both 4-vector potential fields in the second term carry covariant indices;
that is, the $A$-action is euclidian.\cite{Euclidian_footnote} $n(x) =
{\bar\psi}_{\sigma}(x)\,\psi_{\sigma}(x)$ and ${\bm n}_{\text{s}}(x) =
{\bar\psi}_{\sigma_1}(x)\,{\bm\sigma}_{\sigma_1 \sigma_2}\,\psi_{\sigma_2}(x)$
are the electronic number and spin density,\cite{magnetization_footnote}
respectively, with ${\bm\sigma} = (\sigma^x,\sigma^y,\sigma^z)$ the Pauli
matrices, and the third term in Eq.\ (\ref{eq:3.1}) describes the coupling of
the electrons to the electromagnetic field, with ${\bm B} =
{\bm\nabla}\times{\bm A}$ the magnetic induction.\cite{ionic_density_footnote}
Finally, the last term in Eq.\ (\ref{eq:3.1}) describes the spin-orbit
interaction. Note that both terms coupling the scalar potential $\varphi$ to
the fermions carry an extra factor of $i$ compared to what one might expect
from the first quantized Hamiltonian. This has the same origin as the Euclidian
metric mentioned above.\cite{Euclidian_footnote}

Equation (\ref{eq:3.1}) describes a continuum model. Some of the effects we are
interested in are present only in the presence of certain types of lattices,
and we will comment later on the modifications that occur if the electrons are
put on a lattice.

\subsection{Integrating out the photons}
\label{subsec:III.B}

The action, Eq. (\ref{eq:3.1}), depends only bilinearly on the electromagnetic
potential. The latter can therefore be integrated out exactly, albeit at the
expense of creating four-fermion terms. The latter represent electron-electron
interactions that are mediated by the exchange of virtual photons. Technically,
we need the photon propagator, which we can read off the second term in Eq.\
(\ref{eq:3.1}):
\bse
\label{eqs:3.2}
\be
\langle A_{\mu}(x)\,A_{\nu}(y)\rangle = \delta_{\mu\nu}\,{\cal D}(x-y),
\label{eq:3.2a}
\ee
with
\be
{\cal D}(x-y) = -4\pi\,\left(\partial_{\tau}^2/c^2 +
{\bm\nabla}^2\right)^{-1}\,\delta(x-y),
\label{eq:3.2b}
\ee
or, in Fourier space,
\be
{\cal D}(k) = 4\pi/\left(\Omega_n^2/c^2 + {\bm k}^2\right).
\label{eq:3.2c}
\ee
\ese
Here $k \equiv (i\Omega_n,{\bm k})$ comprises a bosonic Matsubara frequency
$\Omega_n = 2\pi T n$ and a wave vector ${\bm k}$.

The result of integrating out the photons exactly is very complicated and
involves interacting electronic modes in both the spin-singlet and spin-triplet
channels, the particle-particle and particle-hole channels, and all
angular-momentum channels. We will restrict ourselves to those terms that are
most relevant for magnetism, i.e. interactions between spin-density
fluctuations, or modes in the $s$-wave particle-hole spin-triplet channel.

\subsubsection{$O(1/c^0)$: Exchange interaction}
\label{subsubsec:III.B.1}

We organize the various contributions to the effective electron-electron
interaction in powers of $1/c$. To zeroth order only the term coupling
$\varphi$ to the number density $n$ in Eq.\ (\ref{eq:3.1}) contributes.
Integrating out $\varphi$ leads to a Coulomb interaction
\bse
\label{eqs:3.3}
\be
S_{\,\text{C}} = -\frac{1}{2} \int dx\,dy\ n(x)\,v_{\text{C}}({\bm x}-{\bm
y})\, \delta(\tau_x - \tau_y)\, n(y),
\label{eq:3.3a}
\ee
with
\be
v_{\,\text{C}}({\bm x}) = e^2/\vert{\bm x}\vert.
\label{eq:3.3b}
\ee
\ese
Here we have neglected the dynamical nature of the photon propagator ${\cal D}$
and have replaced it by its value at $\Omega_n=0$. The reason for this
approximation is that Fermi-liquid effects lead to a dynamical screening of the
Coulomb interaction that is a much larger effect than the relativistic dynamics
inherent in Eqs.\ (\ref{eqs:3.2}).

Equation\ (\ref{eq:3.3a}) contains number density fluctuations at all
wavelengths. If one restricts the theory to interactions between
long-wavelength fluctuations, then this interaction can be rewritten as a sum
of parts that includes an interaction between spin-density fluctuations, see
Ref.\ \onlinecite{AGD_1963} and Appendix \ref{app:B}. The basic point is that
an interaction between number density fluctuations at large wave numbers can be
written as one between spin density fluctuations at small wave numbers. In an
effective low-energy theory that contains only fluctuations at wave numbers
smaller than some cutoff $\lambda$, Eq.\ (\ref{eq:3.3a}) therefore contains a
contribution
\be
S_{\text{ex}} = \frac{\Gamma_{\text{t}}}{2} \int' dx\ {\bm n}_{\text{s}}(x)
\cdot {\bm n}_{\text{s}}(x),
\label{eq:3.4}
\ee
where the prime on the integral indicates that only the small-wave-number
contributions (smaller than $\lambda$) to the spin density ${\bm n}_{\text{s}}$
are to be considered in order to avoid overcounting. As has been explained in
Ref.\ \onlinecite{Belitz_Evers_Kirkpatrick_1998}, it is convenient to choose
the cutoff $\lambda$ as a fixed fraction of the Thomas-Fermi screening wave
number, and the spin-triplet interaction amplitude $\Gamma_{\text{t}}$ is a
Fermi-surface average over $v_{\,\text{C}}({\bm k}-{\bm p})\,\Theta(\vert {\bm
k}-{\bm p}\vert - \lambda)$, with ${\bm k}$ and ${\bm p}$ pinned to the Fermi
surface. The restriction to small wave numbers will be understood from now on,
and we will drop the prime on integrals.

$S_{\text{ex}}$ is the exchange interaction between electronic spin-density
fluctuations that leads ferromagnetism. For later reference we note that
$\Gamma_{\text{t}}$ is dimensionally an energy times a volume which, in this
unrenormalized theory, is on the order of a Rydberg times a Fermi volume, or
$\Gamma_{\text{t}} \approx e^2/\kF^2$.

\subsubsection{$O(1/c^2)$: Dipole-dipole interaction}
\label{subsubsec:III.B.2}

We now turn to terms of $O(1/c^2)$. We first consider the vector potential
${\bm A}$, which couples to the spin density via the ${\bm B}\cdot{\bm
n}_{\text{s}}$ term in Eq.\ (\ref{eq:3.1}). Since the coupling is directly to
the spin channel, no phase space decomposition is necessary and integrating out
${\bm A}$ proceeds as in the classical case, except that the ${\bm
A}$-propagator now is frequency dependent, see Eqs.\ (\ref{eqs:3.2}). We will
comment on the consequences of this frequency dependence in Sec.
\ref{subsubsec:III.C.2} below. Neglecting the dynamical aspects of the
dipole-dipole interaction for now, we obtain a contribution to the effective
action
\bea
S_{\text{d-d}} &=& 2\pi\muB^2 \int dx\ {\bm n}_{\text{s}}(x)\cdot{\bm
n}_{\text{s}}(x) \nonumber\\
&&\hskip 0 pt + \frac{\muB^2}{2} \int dx\,dy\ \delta(\tau_x - \tau_y)\,
n_{\text{s}}^i(x)\,d_{ij}({\bm x}-{\bm y})\, n_{\text{s}}^j(y), \nonumber\\
\label{eq:3.5}
\eea
with $d_{ij}$ from Eq.\ (\ref{eq:2.7b}). The first term has the same form as
the exchange interaction, Eq.\ (\ref{eq:3.4}), but is much smaller, as $\muB^2
\approx \Gamma_{\text{T}}\,(\vF/c)^2$. The second one is the dipole-dipole
interaction between the electron spins; if one replaces the electronic spin
density by its quantum mechanical and thermal average one recovers the
classical dipole-dipole term in Eq.\ (\ref{eq:2.4a}) or (\ref{eq:2.7a}).

\subsubsection{$O(1/c^2)$: Dzyaloshinsky-Moriya and related interactions}
\label{subsubsec:III.B.3}

We now return to the effects of integrating out the scalar potential $\varphi$.
To $O(1/c^2)$ the relevant contribution comes from the cross-term that
multiplies the Coulomb ($\varphi\,n$) term and the spin-orbit (last) term in
Eq.\ (\ref{eq:3.1}). Contracting the two scalar potentials, integrating by
parts, using Eq.\ (\ref{eq:C.2b}), and keeping only terms that are bilinear in
phase-space spin-density fluctuations, we obtain a contribution to the
effective action
\bea
S_{\text{s-o}} &=& \frac{-\muB^2}{2}\, \epsilon_{ilm}\,\epsilon_{ijk} \int
dx\,dy\ {\cal D}(x-y)\, {\bar\psi}_{\sigma_1}(x)\,\sigma^l_{\sigma_1\sigma_4}\,
\nonumber\\
&&\hskip 0 pt \times\left(\frac{\partial}{\partial
y_k}\,\psi_{\sigma_4}(y)\right)\, \left(\frac{\partial}{\partial
y_j}\,{\bar\psi}_{\sigma_3}(y)\right)\, \sigma^m_{\sigma_3 \sigma_2}\,
\psi_{\sigma_2}(x). \nonumber\\
\label{eq:3.6}
\eea
After a Fourier transform, Eq.\ (\ref{eq:3.6}) can be written
\begin{widetext}
\bea
S_{\text{s-o}} &=& \frac{\muB^2}{4}\,\left(\frac{T}{V}\right)^2 \sum_{q,k,p}
{\cal D}({\bm q})\,
\left({\bm\sigma}_{\sigma_1\sigma_4}\times{\bm\sigma}_{\sigma_3\sigma_2}\right)
\cdot \left[({\bm q}\times{\bm p})\, {\bar\psi}_{\sigma_1}(k-q/2)\,
{\bar\psi}_{\sigma_3}(p+q/2)\, \psi_{\sigma_4}(p-q/2)\,
\psi_{\sigma_2}(k+q/2)\right.
\nonumber\\
&&\hskip 150pt \left. - ({\bm q}\times{\bm k})\, {\bar\psi}_{\sigma_1}(k+q/2)\,
{\bar\psi}_{\sigma_3}(p-q/2)\, \psi_{\sigma_4}(p+q/2)\, \psi_{\sigma_2}(k-q/2)
\right].
\label{eq:3.7}
\eea
Here $k = ({\bm k},i\omega_n)$ comprises a wave vector ${\bm k}$ and a
fermionic Matsubara frequency $\omega_n = 2\pi T(n+1/2)$, and $p$ and $q$ are
used analogously. At this point we generalize to an effective interaction
amplitude ${\cal D}_{{\bm k},{\bm p}}({\bm q})$ that depends on ${\bm k}$ and
${\bm p}$ in addition to ${\bm q}$. Such a structure is generated in
perturbation theory from the bare theory, where the interaction amplitude is
simply given by the gauge field propagator, as we demonstrate in Appendix
\ref{app:D}. We then have
\bea
S_{\text{s-o}} &=& \frac{\muB^2}{4}\,\left(\frac{T}{V}\right)^2 \sum_{q,k,p}
\left[{\cal D}_{{\bm k},{\bm p}}({\bm q})\, ({\bm q}\times{\bm p}) + {\cal
D}_{{\bm p},{\bm k}}(-{\bm q})\, ({\bm q}\times{\bm k})\right] \cdot
\left({\bm\sigma}_{\sigma_1 \sigma_4}\times{\bm\sigma}_{\sigma_3
\sigma_2}\right)  \nonumber\\
&&\hskip 100pt \times {\bar\psi}_{\sigma_1}(k-q/2)\,
{\bar\psi}_{\sigma_3}(p+q/2)\, \psi_{\sigma_4}(p-q/2)\, \psi_{\sigma_2}(k+q/2).
\label{eq:3.8}
\eea
Hermiticity requires ${\cal D}^*_{{\bm k},{\bm p}}({\bm q}) = {\cal D}_{{\bm
k},{\bm p}}(-{\bm q})$ (see also Eqs.\ (\ref{eqs:B.3})). We now employ the
phase space decomposition explained in Appendix \ref{app:B} and focus on the
large-angle scattering term, Eq.\ (\ref{eq:B.4b}). Projecting again on the spin
density we obtain
\bse
\label{eqs:3.9}
\be
S_{\text{s-o}} = \frac{\muB^2}{4}\,\frac{T}{V}\sum_q {\bm D}({\bm q}) \cdot
\left({\bm n}_{\text{s}}(q)\times{\bm n}_{\text{s}}(-q)\right),
\label{eq:3.9a}
\ee
where
\bea
{\bm D}({\bm q}) &=& \frac{T}{V} \sum_{{\bm k},{\bm p}} g_{\bm k}\,g_{\bm p}\,
({\bm k}\times{\bm p})\, \left[{\cal D}_{({\bm k}+{\bm p}+{\bm q})/2,({\bm
k}+{\bm p}-{\bm q})/2}({\bm k}-{\bm p}) + {\cal D}_{({\bm k}+{\bm p}-{\bm
q})/2,({\bm k}+{\bm p}+{\bm q})/2}({\bm p}-{\bm k})\right]
\nonumber\\
&&+\frac{T}{2V} \sum_{{\bm k},{\bm p}} g_{\bm k}\,g_{\bm p}\, ({\bm k}-{\bm
p})\times{\bm q}\, \left[{\cal D}_{({\bm k}+{\bm p}+{\bm q})/2,({\bm k}+{\bm
p}-{\bm q})/2}({\bm k}-{\bm p}) - {\cal D}_{({\bm k}+{\bm p}-{\bm q})/2,({\bm
k}+{\bm p}+{\bm q})/2}({\bm p}-{\bm k})\right]
\nonumber\\
&\equiv& {\bm D}^{(1)}({\bm q}) + {\bm D}^{(2)}({\bm q}).
\label{eq:3.9b}
\eea
\ese
\end{widetext}
Here
\be
g_{\bm k} = \epsilonF\,\delta(\epsilon_{\bm k} - \epsilonF)
\label{eq:3.10}
\ee
is a function that results from the projection onto the spin density and pins
${\bm k}$ and ${\bm p}$ to the Fermi surface.\cite{projection_footnote}
Equation (\ref{eq:3.9a}) has the form of the result obtained by
Moriya.\cite{Moriya_1960} In what follows, we discuss the nature of the vector
${\bm D}({\bm q})$ in some more detail.

${\bm D}^{(2)}$ has the form
\bse
\label{eqs:3.11}
\be
{\bm D}^{(2)}({\bm q}) = i\, {\bm q}\times{\bm d}^{\,(2)} + O({\bm q}^3),
\label{eq:3.11a}
\ee
with ${\bm d}^{\,(2)}$ a real vector given by
\bea
{\bm d}^{\,(2)} &=& \frac{T}{2iV} \sum_{{\bm k},{\bm p}} g_{\bm k}\,g_{\bm p}\,
({\bm p}-{\bm k})\, \left[{\cal D}_{({\bm k}+{\bm p})/2,({\bm k}+{\bm
p})/2}({\bm k}-{\bm p})\right.
\nonumber\\
&&\hskip 40pt \left. - {\cal D}_{({\bm k}+{\bm p})/2,({\bm k}+{\bm p})/2}({\bm
p}-{\bm k})\right].
\label{eq:3.11b}
\eea
\ese

${\bm D}^{(1)}$ can be written
\bse
\label{eqs:3.12}
\be
D^{(1)}_i({\bm q}) = i D_{ij}\,q_j + O(q^3),
\label{eq:3.12a}
\ee
with $D_{ij}$ a real rank-2 tensor given by
\be
D_{ij} = \sum_{{\bm k},{\bm p}} g_{\bm k}\,g_{\bm p}\,\epsilon_{ilm}k_l
p_m\,{\cal E}_j({\bm k},{\bm p}),
\label{eq:3.12b}
\ee
where
\bea
{\cal E}_i({\bm k},{\bm p}) &=& \frac{-1}{2} \int d{\bm x}\,d{\bm y}\
(x_i-y_i)\, e^{-i({\bm k}+{\bm p})({\bm x}+{\bm y})/2} \nonumber\\
&&\hskip -20pt \times \int d{\bm z}\ \sin(({\bm k}-{\bm p})\cdot {\bm
z})\,{\cal D}({\bm x},{\bm y};{\bm z})
\label{eq:3.12c}
\eea
\ese
with ${\cal D}({\bm x},{\bm y};{\bm z})$ the Fourier transform of ${\cal
D}_{{\bm k},{\bm p}}({\bm q})$ in analogy to Eq.\ (\ref{eq:B.2b}). $D_{ij}$ has
a symmetric part and an antisymmetric part. The latter can be combined with
${\bm D}^{(2)}$ above to form a contribution to ${\bm D}$ that we denote by
\bse
\label{eqs:3.13}
\be
{\bm D}^{(-)}({\bm q}) = i\,{\bm q}\times{\bm d} + O({\bm q}^3),
\label{eq:3.13a}
\ee
where
\be
d_i = d_i^{\,(2)} + \frac{T}{2iV} \sum_{{\bm k},{\bm p}} g_{\bm k}\,g_{\bm p}\,
(p_i k_j - k_i p_j){\cal E}_j({\bm k},{\bm p}).
\label{eq:3.13b}
\ee
\ese
The symmetric part can be written as a diagonal tensor plus a traceless rank-2
tensor, and the latter can be diagonalized by means of an orthogonal
transformation that amounts to a spatial rotation. The symmetric part of
$D_{ij}$ we thus can write
\bse
\label{eqs:3.14}
\be
D^{(+)}_{ij} = \frac{1}{3}\,\tr D\, \delta_{ij} + a_i\,\delta_{ij}
\label{eq:3.14a}
\ee
({\em no} summation convention) where the $a_i$ obey
\be
\sum_i a_i = 0.
\label{eq:3.14b}
\ee
\ese
If desirable, the $a_i$ can be explicitly constructed from Eq.\
(\ref{eq:3.12b}). We thus have a second contribution to ${\bm D}$ that we
denote by
\be
{\bm D}^{(+)}({\bm q}) = \frac{i}{3}\,\tr D\,{\bm q} + i\left(\begin{array}{c}
a_x q_x \\ a_y q_y \\ a_z q_z\end{array}\right) + O({\bm q}^3),
\label{eq:3.15}
\ee
and
\be
{\bm D}({\bm q}) = {\bm D}^{(+)}({\bm q}) + {\bm D}^{(-)}({\bm q}).
\label{eq:3.16}
\ee

Combining our results, and transforming back to real space, we now have
\bse
\label{eqs:3.17}
\be
S_{\text{s-o}} = S_{\text{DM}} + S_{\text{DM}}' + S_{\text{DM}}'',
\label{eq:3.17a}
\ee
where
\bea
S_{\text{DM}} &=& \frac{\muB^2}{12}\,\tr D \int dx\ {\bm
n}_{\text{s}}(x)\cdot\left({\bm\nabla}\times{\bm n}_{\text{s}}(x)\right),
\nonumber\\
\label{eq:3.17b}\\
S_{\text{DM}}' &=& \frac{\muB^2}{4} \int dx\ {\bm n}_{\text{s}}(x)\cdot
\left[\left(\begin{array}{c} a_x \partial_x \\ a_y
\partial_y \\ a_z \partial_z
\end{array}\right)\times{\bm n}_{\text{s}}(x)\right],
\nonumber\\
\label{eq:3.17c}\\
S_{\text{DM}}'' &=& \frac{\muB^2}{4} \int dx\ \left({\bm\nabla}\cdot{\bm
n}_{\text{s}}(x)\right)\,\left({\bm d}\cdot{\bm n}_{\text{s}}(x)\right).
\label{eq:3.17d}
\eea
\ese
$S_{\text{DM}}$ is the Dzyaloshinsky-Moriya interaction that is believed to be
responsible for the helimagnetism observed in MnSi and
FeGe.\cite{Bak_Jensen_1980} $S_{\text{DM}}'$ is a closely related term but
absent in systems with cubic lattices (e.g., MnSi or FeGe) due to the
constraint, Eq. (\ref{eq:3.14b}). Finally, $S_{\text{DM}}''$ is another term
that is allowed by symmetry, and is generated by the above derivation. All
three of these terms are contained in Moriya's general
result,\cite{Moriya_1960} which takes the form of our Eq.\ (\ref{eq:3.9a})
above, but the effects of $S_{\text{DM}}'$ and $S_{\text{DM}}''$ have, to our
knowledge, not been discussed explicitly.

Note that a necessary condition for any of these interactions to be nonzero is
that the system is not invariant under parity: Both ${\bm D}^{(1)}$ and ${\bm
D}^{(2)}$ can be nonzero only if ${\cal D}_{{\bm k},{\bm p}}({\bm q})$ is odd
under ${\bm q} \to -{\bm q}$, or, equivalently, if ${\cal D}({\bm x},{\bm
y};{\bm z})$ is odd under ${\bm z} \to -{\bm z}$. This implies that the DM
interaction requires a lattice that is not invariant under spatial inversion;
in any continuum model, where the electron-electron interaction is necessarily
even under parity, it vanishes. See Appendix \ref{app:D} and Sec.\ \ref{sec:IV}
for further discussions of this point.

\subsection{Fermionic action, and magnetic order parameter}
\label{subsec:III.C}

We now have the following result. After integrating out the photons, the
effective fermionic action reads
\bea
S_{\text{eff}}[\bar\psi,\psi] &=& S_0 + S_{\text{int}}' + S_{\text{ex}} +
S_{\text{d-d}} + S_{\text{DM}} + S_{\text{DM}}' + S_{\text{DM}}''
\nonumber\\
&\equiv& S_0' + S_{\text{int}}^{\text{t}}.
\label{eq:3.18}
\eea
Here $S_0$ describes non-interacting electrons (either free electrons or band
electrons, depending on the model considered), and $S_{\text{int}}'$ contains
all interactions between modes other than the spin density, which we have not
explicitly considered with the exception of the Coulomb interaction, Eqs.\
(\ref{eqs:3.3}). Collectively we denote these two terms by $S_0'$.
$S_{\text{ex}}$, $S_{\text{d-d}}$, $S_{\text{DM}}$, $S_{\text{DM}}$, and
$S_{\text{DM}}''$ are the exchange, dipole-dipole, and Dzyaloshinsky-Moriya
interactions given by Eqs.\ (\ref{eq:3.4}), (\ref{eq:3.5}), and
(\ref{eqs:3.17}), respectively. $S_{\text{ex}}$ and $S_{\text{d-d}}$ are always
present; which, if any, of the terms in $S_{\text{DM}}$ are nonzero depends on
the details of the lattice structure and absence of spatial inversion symmetry
is a prerequisite for any of them to be nonzero. Collective, we denote the sum
of these interactions in the spin-density or triplet channel by
$S_{\text{int}}^{\text{t}}$.

\subsubsection{Structure of a magnetic order parameter description}
\label{subsubsec:III.C.1}

For applications such as fermionic cold gases one will want to work directly
with the fermionic action. For applications to magnets it is convenient to
introduce a composite field ${\bm M}(x)$ whose expectation value is
proportional to the magnetization. To this end we write
\bse
\label{eqs:3.19}
\be
S_{\text{int}}^{\text{t}} = \frac{1}{2} \int dx\,dy\
n_{\text{s}}^i(x)\,\Gamma_{ij}(x-y)\, n_{\text{s}}^j(y),
\label{eq:3.19a}
\ee
with
\bea
\Gamma_{ij}(x-y) &=& \delta(x-y)\,\Gamma_{\text{t}} + \muB^2\,\delta(\tau_x -
\tau_y)\,d_{ij}({\bm x}-{\bm y}) \nonumber\\
&&\hskip -60pt + \frac{\muB^2}{2}\,\delta(x-y)\,\left[\frac{1}{3}\,\tr
D\,\epsilon_{ikj}\,\partial_k + \frac{1}{2}\,\epsilon_{ikj}\,(a\partial)_k +
\frac{1}{2}\,d_i\,\partial_j\right].
\nonumber\\
\label{eq:3.19b}
\eea
\ese
Here $(a\partial)_k = a_k\,\partial_k$ (no summation convention). We now
decouple Eq.\ (\ref{eq:3.19a}) by means of a Hubbard-Stratonovich
transformation with a bosonic field ${\bm M}(x)$. Neglecting a constant
contribution to the action, this allows us to write
\bea
S_{\text{eff}}[{\bar\psi},\psi,{\bm M}] &=& S_0'[{\bar\psi},\psi]
\nonumber\\
&&\hskip 0pt -\frac{1}{2} \int dx\,dy\ M_i(x)\,\Gamma_{ij}(x-y)\,M_j(y)
\nonumber\\
&&\hskip 0pt + \frac{1}{4} \int dx\,dy\
\left[M_i(x)\,\Gamma_{ij}(x-y)\,n_{\text{s}}^j(y) \right. \nonumber\\
&&\hskip 30pt \left. + n_{\text{s}}^i(x)\, \Gamma_{ij}(x-y)\,M_j(y)\right].
\nonumber\\
\label{eq:3.20}
\eea

If one neglects the interacting part of $S_0'$ this action depends only
bilinearly on the fermion fields, and one can formally integrate out the
fermions in order to obtain a theory entire in terms of the order-parameter
field ${\bm M}$. This is a generalization of, and replaces, the Hertz-Millis
theory.\cite{Hertz_1976, Millis_1993} However, in general this is not a good
strategy since it amounts to integrating out soft excitations, which means that
any order parameter theory will in general not be well behaved. Physically,
these soft quasi-particle excitations can change the nature of the phase
transition,\cite{Belitz_Kirkpatrick_Vojta_2005} or they themselves can become
critical.\cite{Sachdev_2010} In either case they must be treated on equal
footing with the order parameter fluctuations. It therefore is technically
advantageous, and physically more transparent, to work with the coupled field
theory represented by Eq.\ (\ref{eq:3.20}).

\subsubsection{Comments on the magnetization dynamics}
\label{subsubsec:III.C.2}

We conclude this section with a brief discussion of the dynamics of the
dipole-dipole interaction, which we neglected in Sec.\ \ref{subsubsec:III.B.2}.
If we restore the frequency dependence of the photon propagator and expand in
powers of the frequency, then the leading dynamical contribution of the
dipole-dipole interaction to Eq.\ (\ref{eq:3.20}) takes the form
\be
S_{\text{d-d}}^{\text{dyn}} = \muB^2 \frac{T}{V}\sum_k \left(\delta_{ij} -
{\hat k}_i\,{\hat k}_j\right)\, \frac{\Omega_n^2}{c^2{\bm k}^2}\, M_i(k)\,
M_j(-k).
\label{eq:3.21}
\ee
As long as $\Omega_n$ scales as $\vert{\bm k}\vert$, this scales the same as
the $\vert\Omega_n\vert/\vert{\bm k}\vert$ term in the order-parameter theory
that is induced by Fermi-liquid effects,\cite{Hertz_1976} but has a prefactor
that is smaller by a factor of $(\vF/c)^2$. However, in classical dipolar
magnets the order parameter is known to no longer be
conserved.\cite{Frey_Schwabl_1994} That is, $\Omega$ scales as a constant, and
this should be reflected in the quantum theory as well, although it is
currently not known how this  is realized. This suggests that the contribution
shown in Eq.\ (\ref{eq:3.21}) dominates the Fermi-liquid-induced dynamics in a
scaling sense, although it has a small prefactor, and will become important at
sufficiently long time scales.

\section{Discussion, and Conclusion}
\label{sec:IV}

We now discuss the significance of various interactions for a number of
problems.

\subsection{Energy scales, and the significance of the dipole-dipole
interaction}
\label{subsec:IV.A}

As we mentioned in Sec.\ \ref{subsec:III.A}, the energy scale for the exchange
interaction in the bare theory is the atomic scale, or roughly
$100,000\,{\text{K}}$. The corresponding length scale is on the order of
$1\,\AA$. This is not consistent with the experimental fact that magnetic
ordering is observed only at much lower temperatures; e.g., on the order of
$1,000\,{\text{K}}$ in Fe and Ni, on the order of room temperature in FeGe, and
on the order of $30\,{\text{K}}$ in MnSi. The reason for this discrepancy lies
in the fact that the bare theory is renormalized in quantitatively substantial
ways, and the corresponding energy and length scales in the properly
renormalized theory are consistent with experimental
observations.\cite{Lonzarich_Taillefer_1985}

Equations (\ref{eq:3.5}) and (\ref{eqs:3.17}) show that the dipole-dipole and
DM interactions, within the framework of the bare theory, are weaker than the
exchange interaction by a factor of $(\vF/c)^2$, or about
$10^{-4}$.\cite{dipole_strength_footnote} Relative to the bare exchange
interaction, this implies an energy scale on the order of $10\,\text{K}$, which
is comparable with the ordering temperature in MnSi. On the other hand, the
length scale associated with the DM interaction (i.e., the pitch length of the
spin helix\cite{pitch_footnote}) is only about $200\,\AA$ (in
MnSi)\cite{Ishikawa_et_al_1985} to $700\,\AA$ (in
FeGe),\cite{Lebech_Bernhard_Freltoft_1989} or only a factor of $10^{2}$ to
$10^{3}$ larger than the atomic length scale.

These observations indicate that there are strong renormalizations, due to
band-structure and many-body effects, of all terms in the bare action, and that
different terms are renormalized in different ways in different materials.
While this makes it hard to make general statements, the bare theory suggests
that the DM interactions and the dipole-dipole interaction are generically
comparable in strength, and in MnSi, for instance, both are expected to be a
substantial fraction of the (greatly reduced by renormalizations) exchange
interaction. We thus conclude that there is no a priori reason to neglect the
dipole-dipole interaction in any system where the DM interaction is known to be
important. This calls for a re-evaluation of a number of interesting problems,
some of which we list in the following subsection.

\subsection{Significance of the dipole-dipole interaction}
\label{subsec:IV.B}

We conclude by discussing a number of problems where the dipole-dipole
interaction is either known to be important, or might be important, with an
emphasis on low-temperature magnets and other fermion systems.

(1) Classical Heisenberg ferromagnets. This problem was worked on in great
detail by Aharony and Fisher\cite{Aharony_Fisher_1973} for the static critical
behavior and by Frey and Schwabl\cite{Frey_Schwabl_1994} for the dynamical
critical behavior. The renormalization group done by Aharony and Fisher started
from a nonlocal order parameter theory, Eqs.\ (\ref{eqs:2.7}), which leads to a
somewhat nonconventional renormalization procedure. The nonlocality in Eqs.\
(\ref{eqs:2.7}) is due to integrating out the gauge field fluctuations or
photons. It would be interesting to repeat this calculation starting from the
coupled local field theory given by our Eqs.\ (\ref{eqs:2.2}) before these
fluctuations are integrated out.

(2) Classical helimagnets. The standard phase transition treatment for
helimagnetism due to the DM interaction is due to Bak and
Jensen.\cite{Bak_Jensen_1980} Neglecting the dipole-dipole interaction term
they conclude that there is a fluctuation-induced first order phase transition
from paramagnetism to helimagnetism. An interesting question is whether or not
the dipole-dipole interaction modifies this conclusion. This seems especially
relevant for MnSi where the phase transition is at low temperatures, i.e., it
is a weak helimagnet.

(3) It has been shown that in clean itinerant ferromagnets, the ferromagnetic
transition is generically of first order at zero
temperature.\cite{Belitz_Kirkpatrick_Vojta_1999, Belitz_Kirkpatrick_Vojta_2005}
This conclusion ignores the effects of the dipole-dipole interaction terms. It
would be very interesting to investigate if dipolar interactions can modify
this generic conclusion.

(4) Phase ordering is an important problem in ferromagnets.\cite{Bray_1994} The
dipole-dipole interaction terms has not been included in either the classical
or quantum (zero temperature) ferromagnetic phase ordering problems. Simple
arguments indicate it will be important.

(5) Fermionic cold atom systems. Recently there has been a considerable amount
of work on gases of fermions with dipolar interactions, see Ref.\
\onlinecite{Fregoso_Fradkin_2010} and references therein. These systems are
important for fermions in optical lattices. The dipolar interactions also serve
as a mechanism for liquid crystal like phase formation in fermion systems.

(6) The dipole-dipole interaction term is important in the dynamics of
classical antiferromagnets, both in the ordered phase, and near or at the phase
transition if the systems is below its upper critical
dimension.\cite{Frey_Schwabl_1994} Simple considerations suggest that they will
also be important in low-dimensional (1+1 or 2+1) itinerant quantum
antiferromagnets; see Sec.\ \ref{subsubsec:III.C.2} above for one aspect of
this problem.

\appendix

\section{The complete action to $O(1/c^2)$}
\label{app:A}

In this appendix we give the complete action for electrons interacting with
electromagnetic fields in the weakly relativistic limit, up to and including
terms of $O(1/c^2)$. Let $A_{\mu} = (\varphi,-{\bm A})$ be the 4-vector
potential, ${\bm\sigma} = (\sigma_1,\sigma_2,\sigma_3)$ the Pauli matrices, and
$\muB = e/2\me c$ the Bohr magneton. We use standard relativistic notation,
with covariant and contravariant indices related by a Minkowski metric
$g_{\mu\,\nu} = (+,-,-,-)$. Expanding the Dirac equation in powers of $1/c$,
one obtains, to order $1/c^2$, the following Hamiltonian in first
quantization,\cite{Bjorken_Drell_1964, Davydov_1976, higher_order_footnote}
\bse
\label{eqs:A.1}
\be
{\hat H} = {\hat H}_{\text{P}} + {\hat H}_{\text{so}} + {\hat H}_{\text{D}} +
{\hat H}_{\delta m}.
\label{eq:A.1a}
\ee
Here ${\hat H}_{\text{P}}$ is the Pauli Hamiltonian,
\bea
{\hat H}_{\text{P}} &=& \frac{1}{2\me}\,\left(-i{\bm\nabla}-\frac{e}{c}\,{\bm
A}({\bm x},t)\right)^2 + e\,\varphi({\bm x},t) \nonumber\\
&&\hskip 30pt - \muB\,{\bm\sigma}\cdot\left({\bm\nabla}\times{\bm A}({\bm
x},t)\right),
\label{eq:A.1b}
\eea
and
\bea
{\hat H}_{\text{so}} &=& \frac{ie}{4\me^2
c^2}\,{\bm\sigma}\cdot\left({\bm\nabla} \varphi({\bm
x},t)\times{\bm\nabla}\right),
\label{eq:A.1c}\\
{\hat H}_{\text{D}} &=& \frac{-e}{8\me^2 c^2}\,{\bm\nabla}^2 \varphi({\bm
x},t),
\label{eq:A.1d}\\
{\hat H}_{\delta m} &=& \frac{-1}{8\me^2 c^2}\, {\bm\nabla}^4.
\label{eq:A.1e}
\eea
\ese
describe the spin-orbit interaction, the Darwin term, and the relativistic mass
correction, respectively. Via standard techniques,\cite{Negele_Orland_1988,
Kaputsa_1989} this theory can be reformulated in terms of an action that
depends on fermionic (i.e., Grassmann-valued) field $\psi$ and its adjoint
${\bar\psi}$ as well as the $4$-vector-potential field $A_{\mu}$. For the
partition function one obtains
\be
Z = Z_A\,Z_{\text{FP}},
\label{eq:A.2}
\ee
where
\be
Z_A = \int D[{\bar\psi},\psi]\,D[A_{\mu}]\ e^{S_A[{\bar\psi},\psi;A_{\mu}]},
\label{eq:A.3}
\ee
with an action
\bse
\label{eqs:A.4}
\be
S_A[{\bar\psi},\psi;A_{\mu}] = S_{\psi}[{\bar\psi},\psi] + S_A[A_{\mu}] +
S_{\text{c}}[{\bar\psi},\psi;A_{\mu}].
\label{eq:A.4a}
\ee
Here
\bea
S_{\psi}[{\bar\psi},\psi] &=& \int dx\ {\bar\psi}_{\sigma}(x)
\left[-\partial_{\tau} + \frac{1}{2\me}\,{\bm\nabla}^2 + \mu \right.
\nonumber\\
&&\hskip 40pt - \left. \frac{1}{8\me^2
c^2}\,{\bm\nabla}^4\right]\,\psi_{\sigma}(x)
\label{eq:A.4b}
\eea
describes the electrons with chemical potential $\mu$, and
\bea
S_A[A_{\mu}] &=& \frac{1}{8\pi} \int dx\
A_{\mu}(x)\,\left[\frac{1}{c^2}\,\partial_{\tau}^2 +
{\bm\nabla}^2\right]\,A_{\mu}(x) \nonumber\\
&&\hskip -20pt + \frac{1}{8\pi}\,\frac{\rho -1}{\rho} \int dx\
\left[\frac{1}{c}\,\partial_{\tau} + {\bm\nabla}\cdot{\bm A}(x)\right]^2
\label{eq:A.4c}
\eea
describes the electromagnetic fields, with $\rho\in {\cal R}$ a gauge fixing
parameter. We use a 4-vector notation $x \equiv (\tau,{\bm x})$, $\int dx
\equiv \int d\tau\int d{\bm x}$ for space and imaginary time. $S_{\text{c}}$
describes the coupling between the fermions and the electromagnetic field; it
contains four separate contributions:
\be
S_{\text{c}}[{\bar\psi},\psi;A_{\mu}] = S_{\text{c,P}} + S_{\text{c,L}} +
S_{\text{D}} + S_{\text{so}}.
\label{eq:A.4d}
\ee
Here
\bea
S_{\text{c,P}}[{\bar\psi},\psi;A_{\mu}] &=& -ie \int dx\ \varphi(x)\,n(x)
\nonumber\\
&&\hskip 0pt + \muB \int dx\ {\bm B}(x)\cdot{\bm n}_{\text{s}}(x)
\label{eq:A.4e}
\eea
is the Coulomb and Zeeman paramagnetic coupling that is included in the Pauli
equation, with $n(x) = {\bar\psi}_{\sigma}(x)\,\psi_{\sigma}(x)$ and ${\bm
n}_{\text{s}}(x) = {\bar\psi}_{\sigma_1}(x)\,{\bm\sigma}_{\sigma_1
\sigma_2}\,\psi_{\sigma_2}(x)$.
\bea S_{\text{c,L}}[{\bar\psi},\psi;A_{\mu}] &=&
-i2\muB\int dx\
{\bar\psi}_{\sigma}(x)\,{\bm A}(x)\cdot{\bm\nabla}\,\psi_{\sigma}(x) \nonumber\\
&&\hskip 0pt + i\muB\int dx\ ({\bm\nabla}\cdot{\bm A}(x))\,n(x) \nonumber\\
&&\hskip 0pt - \frac{e^2}{2\me c^2} \int dx\ {\bm A}^2(x)\,n(x)
\label{eq:A.4f}
\eea
is the Landau diamagnetic coupling;
\be
S_{\text{D}}[{\bar\psi},\psi;A_{\mu}] = \frac{-ie}{8\me^2 c^2}\int dx\
({\bm\nabla}^2\varphi(x))\,n(x)
\label{eq:A.4g}
\ee
is the Darwin term that, in the relativistic hydrogen atom, leads to the
so-called {\it zitterbewegung\,}, and
\bea
S_{\text{so}}[{\bar\psi},\psi;A_{\mu}] &=& \frac{-e}{4\me^2 c^2} \int dx\
{\bar\psi}_{\sigma_1}(x)\,{\bm\sigma}_{\sigma_1 \sigma_2}
\nonumber\\
&&\hskip 0pt \cdot\,({\bm\nabla}\varphi(x)\times{\bm\nabla})\,
{\psi}_{\sigma_2}(x)
\label{eq:A.4h}
\eea
\ese
is the spin-orbit coupling.

The second factor in Eq.\ (\ref{eq:A.2}) is
\bse
\label{eqs:A.5}
\be
Z_{\text{FP}} = \int D[{\bar\eta},\eta]\ e^{-S_{\text{FP}}[{\bar\eta},\eta]}
\label{eq:A.5a}
\ee
Here $\eta$ is a one-component Grassmannian field known as a Fadeev-Popov ghost
field, with ${\bar\eta}$ its adjoint, that is governed by an action
\be
S_{\text{FP}}[{\bar\eta},\eta] = \int dx\
{\bar\eta}(x)\,\partial_{\mu}\,\partial^{\mu}\,\eta(x).
\label{eq:A.5b}
\ee
\ese

\section{Phase-space decomposition of interaction terms}
\label{app:B}

For completeness, in this appendix we briefly recapitulate the arguments that
lead to the generation of a spin-spin interaction, Eq.\ (\ref{eq:3.4}), from a
density-density interaction, Eqs.\ (\ref{eqs:3.3}). For further discussion, see
Refs.\ \onlinecite{AGD_1963}, and \onlinecite{Belitz_Kirkpatrick_1997},
\onlinecite{Belitz_Evers_Kirkpatrick_1998}; the latter also explain the
relation to the work by Shankar.\cite{Shankar_1994}

Consider an electron-electron interaction with an interaction amplitude $W$.
For simplicity we assume that the interaction is purely static, and
translationally invariant, but otherwise general. The action has the form
\begin{widetext}
\bea
S &=& \frac{-1}{2} \int d{\bm x}_1\,\ldots\, d{\bm x}_4 \int d\tau\ W({\bm x}_1
-{\bm x}_2, {\bm x}_3 - {\bm x}_4; ({\bm x}_3 + {\bm x}_4 - {\bm x}_1 - {\bm
x}_2)/2) \,\tau_{\sigma_1\,\sigma_2\,\sigma_3\,\sigma_4}\,
{\bar\psi}_{\sigma_1}({\bm x}_1,\tau)\,{\bar\psi}_{\sigma_3}({\bm x}_3,\tau)\,
\nonumber\\
&&\hskip 300pt \times \psi_{\sigma_4}({\bm x}_4,\tau)\,\psi_{\sigma_2}({\bm
x}_2,\tau),
\label{eq:B.1}
\eea
with $\tau$ a general rank-4 tensor. We define Fourier transforms
\bse
\label{eqs:B.2}
\bea
{\bar\psi}_{\sigma}(k) = \sqrt{T/V} \sum_k e^{-ikx}\,
{\bar\psi}_{\sigma}(k)\qquad &,&\qquad \psi_{\sigma}(k) = \sqrt{T/V} \sum_k
e^{ikx}\,\psi_{\sigma}(k),
\label{eq:B.2a}\\
W_{{\bm k},{\bm p}}({\bm q}) &=& \int d{\bm x}\,d{\bm y}\,d{\bm z}\ W({\bm
x},{\bm y};{\bm z}), \label{eq:B.2b}
\eea
\ese
with $kx = {\bm k}\cdot{\bm x} - \omega_n \tau$ where ${\bm k}$ is a wave
vector and $\omega_n = 2\pi T (n+1/2)$ a fermionic Matsubara frequency.
Hermiticity requires
\bse
\label{eqs:B.3}
\bea
W^*_{{\bm k},{\bm p}}({\bm q}) &=& W_{{\bm k},{\bm p}}(-{\bm q}),
\label{eq:B.3a}\\
\tau^*_{\sigma_1 \sigma_2 \sigma_3 \sigma_4} &=& \tau_{\sigma_2 \sigma_1
\sigma_4 \sigma_3}.
\eea
\ese
We then have
\bse
\label{eqs:B.4}
\bea
S &=& \frac{-1}{2}\,\left(\frac{T}{V}\right)^2 \sum_{{\bm k},{\bm p},{\bm q}}
W_{{\bm k},{\bm p}}({\bm q})\ \tau_{\sigma_1\,\sigma_2\,\sigma_3\,\sigma_4}\
{\bar\psi}_{\sigma_1}(k-q/2)\,{\bar\psi}_{\sigma_3}(p+q/2)\,\psi_{\sigma_4}(p-q/2)\,
\psi_{\sigma_2}(k+q/2)
\label{eq:B.4a}\\
&=& \frac{-1}{2}\,\left(\frac{T}{V}\right)^2 \sum_{{\bm k},{\bm p},{\bm q}}
W_{({\bm k}+{\bm p}-{\bm q})/2,({\bm k}+{\bm p}+{\bm q})/2}({\bm p}-{\bm k})\
\tau_{\sigma_1\,\sigma_2\,\sigma_3\,\sigma_4}\
{\bar\psi}_{\sigma_1}(k-q/2)\,{\bar\psi}_{\sigma_3}(p+q/2)\, \psi_{\sigma_4}(k+q/2)\,
\psi_{\sigma_2}(p-q/2) \nonumber\\
\label{eq:B.4b}\\
&=& \frac{-1}{2}\,\left(\frac{T}{V}\right)^2 \sum_{{\bm k},{\bm p},{\bm q}}
W_{({\bm p}-{\bm k}+{\bm q})/2,({\bm k}-{\bm p}+{\bm q})/2}({\bm p}+{\bm k})\
\tau_{\sigma_1\,\sigma_2\,\sigma_3\,\sigma_4}\ {\bar\psi}_{\sigma_1}(-k+q/2)\,
{\bar\psi}_{\sigma_3}(k+q/2)\,
\nonumber\\
&&\hskip 300pt \times \psi_{\sigma_4}(-p+q/2)\,\psi_{\sigma_2}(p+q/2).
\label{eq:B.4c}
\eea
\ese
\end{widetext}
As long as all wave vectors are summed over, all three of these expressions are
identical. If one restricts the summation in such a way that both $\vert{\bm
q}\vert$ and the modulus of the third argument of $W$ are smaller than a cutoff
wave number $\lambda$, then we can represent the action as a sum of all three
terms. They represent small-angle scattering, large-angle scattering, and
$2\kF$-scattering, respectively. Alternatively, if one is interested in only
one of these channels, one can pick the appropriate formulation of $S$,
restrict oneself to small wave numbers, and neglect the other channels. For our
purposes, we are interested in the large-angle scattering channel, Eq.\
(\ref{eq:B.4b}). By choosing $\tau_{\sigma_1\,\sigma_2\,\sigma_3\,\sigma_4} =
\sigma^0_{\sigma_1\,\sigma_2}\, \sigma^0_{\sigma_3\,\sigma_4}$ and making use
of Eq.\ (\ref{eq:C.2a}) we obtain a term that has the structure of the exchange
interaction, Eq.\ (\ref{eq:3.4}) (in addition to a contribution to the
number-density interaction). The resulting spin-triplet mode is more
complicated than a pure spin density, but it has an overlap with the spin
density and can be restricted to the latter by the projection technique
explained in Ref.\ \onlinecite{Belitz_Kirkpatrick_1997}. Note that a repulsive
Coulomb interaction results in an attractive exchange interaction due to a
commutation of fermion fields that is necessary to write the result in the form
of Eq.\ (\ref{eq:3.4}).

Similarly, by choosing $\tau_{\sigma_1\,\sigma_2\,\sigma_3\,\sigma_4} =
\sigma^0_{\sigma_1\,\sigma_2}\, {\bm\sigma}_{\sigma_3\,\sigma_4}$ and making
use of Eq.\ (\ref{eq:C.2b}) we obtain the structure found in Sec.\
\ref{subsubsec:III.B.3} (in addition to terms that couple the number density
and the spin density).

\section{Properties of Pauli matrices}
\label{app:C}

Here we give some properties of the Pauli matrices that were used in Sec.\
\ref{sec:III}. Let ${\bm\sigma} = (\sigma^x,\sigma^y,\sigma^z) \equiv
(\sigma^1,\sigma^2,\sigma^3)$ be the Pauli matrices with the commutator
property

\be
\epsilon_{ijk}\,\sigma^i\,\sigma^j = i\sigma^k \qquad (i,j,k = 1,2,3),
\label{eq:C.1}
\ee
and $\sigma^0$ the $2\times 2$ unit matrix. Then the following identities hold:
\bse
\label{eqs:C.2}
\bea
\sigma^0_{\sigma_1 \sigma_2}\, \sigma^0_{\sigma_3 \sigma_4} &=& \frac{1}{2}\,
\sigma^0_{\sigma_1 \sigma_4}\, \sigma^0_{\sigma_3 \sigma_2} + \frac{1}{2}\,
{\bm\sigma}_{\sigma_1 \sigma_4}\cdot{\bm\sigma}_{\sigma_3 \sigma_2},\qquad
\label{eq:C.2a}\\
\sigma^0_{\sigma_1 \sigma_2}\, {\bm\sigma}_{\sigma_3 \sigma_4} &=&
\frac{1}{2}\, {\bm\sigma}_{\sigma_1 \sigma_4}\, \sigma^0_{\sigma_3 \sigma_2} +
\frac{1}{2}\, \sigma^0_{\sigma_1 \sigma_4}\, {\bm\sigma}_{\sigma_3 \sigma_2}
\nonumber\\
&&\hskip 20pt + \frac{i}{2}\, {\bm\sigma}_{\sigma_1
\sigma_4}\times{\bm\sigma}_{\sigma_3 \sigma_2}.
\label{eq:C.2b}
\eea
\ese
Equation (\ref{eq:C.2a}) is easily checked by a direct calculation, and Eq.\
(\ref{eq:C.2b}) follows by multiplying Eq.\ (\ref{eq:C.2a}) by ${\bm\sigma}$
from the right and using Eq.\ (\ref{eq:C.1}).

\section{Nonlocal electron-electron interaction}
\label{app:D}

In this appendix we demonstrate that a general interaction amplitude of the
type used in Eq.\ (\ref{eq:B.1}) is generated in perturbation theory from an
ordinary two-body interaction. For simplicity, we consider spinless fermions
interacting via an interaction potential $V$; spin dependence or gradients, as
in the last term of Eq.\ (\ref{eq:3.1}), are easily added. We also consider a
translationally invariant model; we will comment on this feature below.

\begin{figure}[t]
\vskip -0mm
\includegraphics[width=6.0cm]{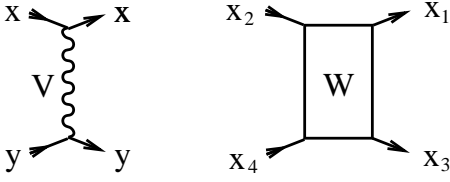}
\caption{Bare interaction $V$ and effective interaction $W$.}
\label{fig:D.1}
\end{figure}

\begin{figure}[t,h]
\vskip -0mm
\includegraphics[width=6.0 cm]{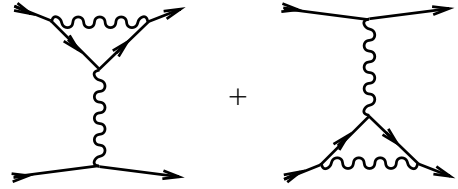}
\caption{A second-order contribution to $W$.}
\label{fig:D.2}
\end{figure}

The bare electron-electron interaction is described by a term in the action
\be
S = \int dx\,dy\ V(x-y)\,{\bar\psi}(x)\,{\bar\psi}(y)\,\psi(y)\,\psi(x).
\label{eq:D.1}
\ee
Our goal is to construct an effective interaction of the form
\bea
S_{\text{eff}}\!\! &=&\!\! \int dx_1 \ldots dx_4\, \nonumber\\
&&\hskip 0pt \times W(x_1\!-\!x_2,x_3\!-\!x_4;(x_3\!+\!x_4\!-\!x_1\!-\!x_2)/2)
\nonumber\\
&&\hskip 0pt \times {\bar\psi}(x_1)\, {\bar\psi}(x_3)\,\psi(x_4)\, \psi(x_2)
\label{eq:D.2}
\eea
that has a structure necessary for contributing to the tensor $D_{ij}$, Eqs.\
(\ref{eq:3.12b}, \ref{eq:3.12c}), see Fig.\ \ref{fig:D.1}.
The bare interaction corresponds to
\be
W^{(1)}(x,y,;z) = \delta(x)\,\delta(y)\,V(z).
\label{eq:D.3}
\ee
This does not contribute to $D_{ij}$ since it enforces $x=y=0$. More generally,
contributions to $W$ with the property $W(x,y,;z) = W(y,x;z)$ do not contribute
to $D_{ij}$. At second order in $V$, there are several diagrams, both
tree-level and one-loop diagrams, that have this property. Consider, however,
the diagrams shown in Fig.\ \ref{fig:D.2}.
They correspond to
\begin{widetext}
\bea
W^{(2)}(x,y;z) &=& \delta(x)\,V(-y) \int dx'\ G(-x'+y/2)\, G(x'+y/2)\,
V(-x'+z/2) \nonumber\\
&&\hskip 0pt + \delta(y)\,V(-x)\ \int dx'\ G(-x'+x/2)\, G(x'+x/2)\, V(-x'+z/2),
\label{eq:D.4}
\eea
\end{widetext}
where $G(x-y) = \langle\psi(x){\bar\psi}(y)\rangle$ is the electron propagator.

In order for $W$ to contribute to $D_{ij}$ we must have $W(x,y;-z) =
-W(x,y;z)$, see Eq.\ (\ref{eq:3.12c}). From Eq.\ (\ref{eq:D.4}) we see that
this is the case if and only if $V(x)$ has an odd component $V^-(-x) = -V(x)$.
As far as the contribution to $D_{ij}$ is concerned we can thus replace $V$ in
Eq.\ (\ref{eq:D.4}) by $V^-$, and this automatically ensures $W(y,x;z) =
-W(x,y;z)$. If $V$ is the propagator of a scalar field, such as the quantity
${\cal D}$ in Sec.\ \ref{sec:III}, then these symmetry properties can obviously
not be realized. However, in a realistic solid-state model $V$ represents the
screened Coulomb interaction, $V(x,y) = v_{\text{C}}({\bm x}-{\bm
y})/\epsilon({\bm x},{\bm y})$, and the dielectric function $\epsilon$ has a
contribution from the lattice in addition to an electronic contribution. The
latter will only have the symmetry of the space group of the lattice, and on a
lattice without inversion symmetry one will have $\epsilon({\bm y},{\bm x})
\neq \epsilon({\bm x},{\bm y})$. Translational symmetry will also be broken, of
course, but this is not necessary for making the DM interaction nonzero, as the
above example shows. Note that one can consider a coarse-grained continuum
theory for the DM interaction, see Eq.\ (\ref{eq:2.8}), but the relevant Landau
coefficient {\em must} depend on the underlying lattice and vanish if the true
continuum limit is taken.

\acknowledgments

We would like to thank Achim Rosch and Thomas Vojta for discussions. This
research was initiated at the Aspen Center for Physics, and supported by the
National Science Foundation under Grant Nos. DMR-09-29966 and DMR-09-01907.

%\bibliography{dipole_derivation}

\begin{thebibliography}{36}
\bibitem[{mag()}]{magnetization_footnote}
\bibinfo{note}{We define the classical magnetization ${\bm M}$ as the (quantum
  mechanical and thermal) average $\langle{\bar\psi}\,{\bm\sigma}\,
  \psi\rangle$ (see Sec.\ \ref{sec:III} below), so a $g$-factor $g=2$ is
  already included in the definition of ${\bm M}$.}
\bibitem[{ion()}]{ionic_density_footnote}
\bibinfo{note}{The scalar potential couples to the ionic density in addition to
  the electronic one. This coupling, which leads to important qualitative
  effects in lattice systems, is neglected in the explicit development of the
  theory in this section, but will be taken into later, see Sec.\
  \ref{subsubsec:III.B.3} and Appendix \ref{app:D}.}
\bibitem[{pro()}]{projection_footnote}
\bibinfo{note}{$g_{\bm k}$ also depends on ${\bm q}$, but its leading ${\bm
  q}$-dependence is quadratic and thus contributes only to terms in the action
  that are cubic in the gradient operator.}
\bibitem[{hig()}]{higher_order_footnote}
\bibinfo{note}{In the usual expansion of the Dirac equation, some additional
  term naturally arise and are often included in the spin-orbit interaction.
  Here we keep only terms that are of $O(1/c^2)$.}
\end{thebibliography}

\end{document}